\documentstyle[12pt,epsfig]{article} 
\begin{document}
\title{Coupling gravitomagnetism--spin and Berry's phase}
\author{ A. Camacho
\thanks{email: acamacho@nuclear.inin.mx} \\
Department of Physics, \\
Instituto Nacional de Investigaciones Nucleares\\
Apartado Postal 18--1027, M\'exico, D. F., M\'exico.}
\date{}
\maketitle

\begin{abstract}
Resorting to Berry's phase, a new idea to detect, at quantum level, the gravitomagnetic field of any metric theory of gravity, is put forward. It is found in this proposal that the magnitude of the gravitomagnetic field appears only in the definition of the adiabatic regime, but not in the magnitude of the emerging geometric phase. In other words, the physical parameter to be observed does not involve, in a direct way, (as in the usual proposals) the tiny magnitude of the gravitomagnetic field.  
\end{abstract}
\bigskip
\bigskip

\section{Introduction}
\bigskip

One of the effects predicted by many metric theories of gravity [1], among them general relativity (GR) [2], which has no Newtonian counterpart, is the so called gravitomagnetic field, that emerges as a consequence of mass--energy currents. Though this field has already been detected [3], it has to be clearly stated that this experiment was performed employing classical systems. Nevertheless, the possible consequences on quantum systems, particularly on the coupling spin--gravitomagnetic field, has always been forgotten, i.e., it is always assumed that the coupling orbital angular momentum--gravitomagnetism can be extended to explain the coupling spin--gravitomagnetic field [4]. Neverwithstanding this assumption must be subject to experimental scrutiny [5]. 

One of the problems in the detection of gravitomagnetism comprises the fact that it involves tiny perturbations in the orbit  
of the used satellites [3]. At this point we may pose the following question: could this field be detected without having to measure very small changes, either in the trajectory or in other physical observable? 

Though we could immediately say {\it no} as answer to this question, there is a possible way to beat this rap. Indeed, if we resort to Berry's phase [6] we may notice that it is possible to have geometric phase changes, which in the interaction of a rotating magnetic field and a quantum particle with spin, depend, exclusively, upon the solid angle that the magnetic field sweeps out, and not on the strength of the field.

Hence from this last remark, in order to give a {\it yes} as answer to our question, we will consider in this work a $1/2$ spin particle immersed in the gravitomagnetic field of a rotating sphere (this field will be described in the PPN formalism, and we will consider the case of any metric theory of gravity), where its rotation axis also spins (sufficiently slow). It will be proved that the interaction spin--gravitomagnetism predicts a geometric phase for the wave function, the one does not depend upon the strength of the interaction. Additionally, it will be shown that the condition that defines the adiabatic regime does involve the strength of our field.

In other words, we put forward the following experiment: a beam of $1/2$ spin particles (all in the initial state) is split in two, one of the beams will not be allowed to interact with $\vec {J}$ (this can be done taking this beam far away from the sphere), whereas the second one will have its spin state pointing always in the direction of $\vec {J}$ (this is obtained, as a consequence of the adiabatic theorem, when $\vec {J}$ spins, sufficiently slow, around a certain axis). After this angular momentum vector completes one cycle we recombine our two beams. The final probability will contain a geometric phase factor, which shall be nonvanishing if the coupling spin--gravitomagnetism is not the trivial one.
\bigskip
\bigskip

\section{Berry's phase and gravitomagnetism}
\bigskip

Let us consider a rotating uncharged, idealized spherical body with mass $M$ and angular momentum $\vec {J}$. In the weak field and slow motion limit the gravitomagnetic field may be written, using the PPN parameters $\Delta_1$ and $\Delta_2$ [7], as

{\setlength\arraycolsep{2pt}\begin{eqnarray}
\vec {B} = \Bigl({7\Delta_1 + \Delta_2\over 4}\Bigr){G\over c^2}{\vec {J} - 3(\vec {J}\cdot\hat {x})\hat {x}\over |\vec {x}|^3}.
\end{eqnarray}}
\bigskip

Here ${7\Delta_1 + \Delta_2\over 4} = 2$ implies GR, while Brans--Dicke (BD) appears if ${7\Delta_1 + \Delta_2\over 4} = {12 + 8\omega\over 8 + 4\omega}$. An interesting point emerges in Ni's theory [8], where ${7\Delta_1 + \Delta_2\over 4} = 0$, i.e., there is no gravitomagnetic field. 

Additionally we will assume that $\vec {J}$ rotates around a certain axis, $\vec {e}_3$, with angular velocity $\omega$, and that the direction of this axis and that of the angular momentum defines an angle $\theta$. In other words, in our coordinate system 
{\setlength\arraycolsep{2pt}\begin{eqnarray}
\vec {J} = J\Bigl[\cos(\omega t)\sin(\theta)\vec {e}_1 + \sin(\omega t)\sin(\theta)\vec {e}_2 +  \cos(\theta)\vec {e}_3\Bigr].
\end{eqnarray}}
\bigskip

Let us consider a spin $1/2$ system immersed in the gravitomagnetic field of $M$, and located on $\vec {e}_3$ at a distance $r$ from the center of the sphere. A usual [4], we assume that the expression that describes the precession of orbital angular momentum can be also used for the description of the dynamics in the case of intrinsic spin, though we must underline the fact that up to now there is a lack of experimental evidence in this direction [9]. Therefore, the Hamiltonian reads

{\setlength\arraycolsep{2pt}\begin{eqnarray}
H = - \vec {S}\cdot\vec {B}.
\end{eqnarray}}
\bigskip

Defining

{\setlength\arraycolsep{2pt}\begin{eqnarray}
\omega_1= {7\Delta_1 + \Delta_2\over 2}{GJ\over c^2r^3},
\end{eqnarray}}
\bigskip

we may rewrite (3) as
{\setlength\arraycolsep{2pt}\begin{eqnarray}
H = - {\hbar\omega_1\over 2}
\left(\begin{array}{cc}
      -2\cos(\theta), & e^{-i\omega t}\sin(\theta) \\
       e^{i\omega t}\sin(\theta), & 2\cos(\theta)
       \end{array}\right).
\end{eqnarray}}

The energy eigenvalues are

{\setlength\arraycolsep{2pt}\begin{eqnarray}
E_{(\pm)} = \pm{\hbar\omega_1\over 2}\sqrt{1 + 3\cos^2(\theta)}.
\end{eqnarray}}
\bigskip

The eigenvector associated with $E_{(+)}$ reads

{\setlength\arraycolsep{2pt}\begin{eqnarray}
\psi_{(+)}(t) = {\sin(\theta)\over\sqrt{2 + 6\cos^2(\theta) -4\cos(\theta)\sqrt{1 + 3\cos^2(\theta)}}}
\left(\begin{array}{c}
      1 \\
      {2\cos(\theta)- \sqrt{1 + 3\cos^2(\theta)}\over\sin(\theta)}e^{i\omega t}
       \end{array}\right).
\end{eqnarray}}

According to Berry [6], if $\omega_1 >>\omega$, and the initial spin state is $\psi_{(+)}(t = 0)$, then the spin state is given by

{\setlength\arraycolsep{2pt}\begin{eqnarray}
\Psi_{(+)}(t) = e^{iE_{(+)}t/\hbar}e^{i\gamma_{(+)}(t)}\psi_{(+)}(t), 
\end{eqnarray}}

where $\gamma_{(+)}(t)$ is Berry's phase, a geometric term given by [6]

{\setlength\arraycolsep{2pt}\begin{eqnarray}
\gamma_{(+)}(t) = i\int_0^t<\psi_{(+)}(t')\vert {\partial\psi_{(+)}(t')\over\partial t'}>dt'.
\end{eqnarray}}

With (7) we obtain

{\setlength\arraycolsep{2pt}\begin{eqnarray}
\gamma_{(+)}(t) = -\omega t\Bigl[1 - {\sin^2(\theta)\over 2 + 6\cos^2(\theta) -4\cos(\theta)\sqrt{1 + 3\cos^2(\theta)}}\Bigr].
\end{eqnarray}}

It is readily seen that this phase is independent of the magnitude of the gravitomagnetic field.

In the case $t = {2\pi\over \omega}$ (which means that $\vec {J}$ has completed one rotation around $\vec {e}_3$) Berry's phase reads

{\setlength\arraycolsep{2pt}\begin{eqnarray}
\gamma_{(+)}(t) = -2\pi\Bigl[1 - {\sin^2(\theta)\over 2 + 6\cos^2(\theta) -4\cos(\theta)\sqrt{1 + 3\cos^2(\theta)}}\Bigr].
\end{eqnarray}}
\bigskip
\bigskip

The adiabatic regime appears if $\omega_1 >>\omega$, and we may rephrase this condition, employing (4), as

{\setlength\arraycolsep{2pt}\begin{eqnarray}
{7\Delta_1 + \Delta_2\over 2}{GJ\over c^2r^3} >> \omega.
\end{eqnarray}}
\bigskip
\bigskip

\section{Conclusions}
\bigskip

A $1/2$ spin particle immersed in the gravitomagnetic field of a rotating sphere (whose rotation axis also spins, sufficiently slow), has been considered. It was proved that the interaction spin--gravitomagnetism predicts a geometric phase for the wave function (Berry's phase), the one does not depend upon the strength of the interaction. This phase is a function of the angular velocity $\omega$, and of the elapsed time. In other words, the physical parameter to be observed does not involve, in a direct way, the magnitude of the gravitomagnetic field. In this sense, this proposal is quite different to the usual experimental ideas, which must detect tiny changes in physical parameter [2, 3]. Furthermore, the present approach could allow us to confront, against measurement readouts, the usual assumption concerning the coupling gravitomagnetism--spin, an issue that lacks experimental support [9].

Additionally, it was shown that the condition that defines the adiabatic regime involves the strength of the field. If we assume, for the sake of simplicity, that our sphere is a homogeneous one (which implies $J = 2MR^2\Omega/5$, here $\Omega$ is the angular velocity of $M$), then (12) renders an inequality, for the experimental parameters, that entails the validity of the adiabatic regime.

{\setlength\arraycolsep{2pt}\begin{eqnarray}
{MR^2\Omega\over\omega r^3} >> {5c^2\over G(7\Delta_1 + \Delta_2)}.
\end{eqnarray}}
\bigskip

Summing up, we have put forward the following experiment: a beam of $1/2$ spin particles (all in the initial state) is split in two, one of the beams will not be allowed to interact with $\vec {J}$, whereas the second one will have its spin state pointing always in the direction of $\vec {J}$. After the angular momentum vector completes one cycle we recombine the two beams. If $\Phi_0$ denotes the first beam, then the final probability will look like $\vert\Phi\vert = \vert\Phi_0\vert\cos^2(\zeta /2)$, where $\zeta$ is a phase factor that shall depend upon Berry's phase, though it must contain also a dynamic phase (this last contribution reads ${\pi\omega_1\over\omega}\sqrt{1 + 3\cos^2(\theta)}$).
\bigskip
\bigskip

\Large{\bf Acknowledgments.}\normalsize
\bigskip

The author would like to thank A. A. Cuevas--Sosa for his 
help. This work was partially supported by CONACYT (M\'exico) Grant No. I35612--E.

\end{document}